\documentstyle[12pt,flushrt]{article}

\newcommand{\be}{\begin{equation}}
\newcommand{\ee}{\end{equation}}
\newcommand{\mpl}{ {M_{\rm pl}}} 
\newcommand{\mbh}{ {M_{\rm bh}}} 
\newcommand{\mbhc}{ {M_{\rm bh} \ast}} 
\newcommand{\gp}{ {\Gamma_P}} 
\newcommand\pp{\parshape 2 0.0truecm 14.25truecm 2truecm 12.25truecm}

\begin{document}

\baselineskip 18pt 

\centerline{\bf THE GRAVITATIONAL DEMISE OF COLD DEGENERATE STARS} 
\bigskip 
\centerline{\bf Fred C. Adams, Gregory Laughlin, and Manasse Mbonye} 
\bigskip 
\centerline{Physics Dept., University of Michigan, Ann Arbor, MI 48109, USA}
\bigskip 
\centerline{\bf Malcolm J. Perry} 
\bigskip 
\centerline{DAMTP, University of Cambridge, Silver Street, Cambridge, 
CB3 9EW, England} 
\bigskip 
\centerline{\it 19 May 1998} 
\bigskip 

\begin{abstract} 
We consider the long term fate and evolution of cold degenerate stars 
under the action of gravity alone. Although such stars cannot emit 
radiation through the Hawking mechanism, the wave function of the star 
will contain a small admixture of black hole states.  These black hole 
states will emit radiation and hence the star can lose its mass energy 
in the long term.  We discuss the allowed range of possible degenerate 
stellar evolution within this framework.  
\end{abstract}


\newpage 
\bigskip 
\medskip 
\centerline{\bf I. INTRODUCTION}
\bigskip 

White dwarfs are usually considered to be the final state of stellar
evolution for nearly all stars in the universe [1]. The remaining
stars are destined to end their lives as neutron stars or black holes.
A fundamental astrophysical question is to determine the long term
fate and evolution of these stellar remnants. If the proton is
unstable through some process at the unification scale $M_X$ [2],
then long term stellar evolution will be governed by the physics of
proton decay with a time scale of 
\be 
\tau_P \sim 10^{32} {\rm yr} (M_X/10^{15} {\rm GeV} )^4 \, . 
\label{eq:gut}
\ee
This evolutionary path has been studied previously [1, 3, 4]. In
this paper, we consider the ultimate fate of these cold degenerate
stars in the absence of unification scale proton decay, i.e., due to
the influence of only the gravitational force.  In particular, we
consider processes involving black holes and their effects on the long
term evolution of these degenerate stars.

If the proton were truly stable, then nature presents us with a
curious state of affairs. Black holes of stellar mass, which are much
more tightly bound than degenerate stars, will evaporate through the
Hawking effect [5] with a lifetime of only $\sim 10^{66}$ yr [6--8]. 
Although this time scale seems long compared to the current age
of the universe ($\sim 10^{10}$ yr), it is vastly shorter than the
expected lifetimes of white dwarfs and neutron stars.  These stellar
remnants are essentially zero temperature objects; they exist in
hydrostatic equilibrium due to quantum mechanical degeneracy pressure
[9].  In the absence of proton decay, these remnants would live
almost forever [1, 10].  In this case, black holes would evaporate
almost instantaneously compared to white dwarfs and neutron stars.

Since degenerate stars are far too extended to have event horizons,
they cannot emit Hawking radiation in the usual manner available to
black holes [11] (see also [6, 8, 12, 13]).  Ref. [11] proved that
in a globally static spacetime, such as that produced by a degenerate
star, there is no particle creation.  These results break down in
black hole spacetimes because they are not globally static, where
static means both time independent and invariant under time reversal.
For black holes, the staticity breaks down at and beyond the event
horizon.

After cooling to essentially zero temperature, degenerate stars can
only lose energy through processes that utilize black holes, which
{\it can} radiate. This avenue is studied in this present paper.
Notice that this approach implicitly includes energy loss from proton
decay via gravitation [14--17], since such processes involve
microscopic black holes.

Motivated by the above discussion, we write the wave function of 
a cold degenerate star in the suggestive form 
\be 
| {\rm star} \rangle = \cos \theta | N_\star \rangle + 
\sin \theta | BH \rangle \, , 
\ee
where $| N_\star \rangle$ represents the usual stellar configurations 
(superpositions of quantum mechanical states of $N_\star$ = few 
$\times 10^{57}$ particles) and $|BH\rangle$ represents superpositions 
of the possible black hole states.  The total probability of the star 
being in a black hole state or containing a black hole is thus 
${\cal P} = \sin^2 \theta \approx \theta^2 \ll 1$.

In general, the black hole part of the wave function will contain 
many different contributions, i.e., 
\be 
| BH \rangle = \sum_{j} A_j | \mbh \rangle \, , 
\ee
where $| \mbh \rangle$ represents the wave function for a black hole 
of mass $\mbh$ and must be a superposition of many states. The quantity 
$|A_j|^2$ is the corresponding probability of the star being 
in a black hole state. The total probability is thus given by 
\be 
{\cal P} = \sin^2 \theta = \sum_{j}  |A_j|^2 \, .  
\ee
We can estimate the probabilities $P_j = |A_j|^2$ from several different 
mechanisms, as discussed below. 

\bigskip 
\centerline{\bf II. BLACK HOLE PROCESSES IN STARS} 
\bigskip 

\bigskip 
\noindent
{\bf A. Dyson Tunneling} 

On sufficiently long time scales, white dwarfs and neutron stars 
can experience tunneling events in which the star, or portions of 
the star, tunnel through the potential energy barrier produced by 
degeneracy pressure [10].  This process can, in principle, result 
in the formation of black holes.  

We first consider the probability of tunneling through the barrier 
provided by degeneracy pressure.  Such tunneling is a necessary 
but not sufficient condition for black hole formation. 
The time scales for tunneling processes are given by the
standard formula,
\be
\tau = \tau_0 \, {\rm e}^{S_T} \, , 
\ee
where $\tau_0$ is the natural oscillation time of the system.
The action integral $S_T$ can be written
\be
S_T = {2 \over \hbar} \, \int \, \bigl[ 2 M V(r) \bigr]^{1/2}
\, dr \, , 
\label{eq:action} 
\ee
where $V(r)$ represents the height of the potential energy barrier 
as a function of the radial coordinate $r$. 

For a white dwarf supported by degenerate electrons, the potential
energy can be obtained from integrals over the distribution functions
for fermions [7]. Following standard formalism, we define
$x=p_F/m_e$, where $p_F$ is the Fermi momentum and $m_e$ is the
electron mass. The number density of electrons is then given by $n_e$
= $x^3 m_e^3 / 3 \pi^2$ (we take $\hbar=1=c$).  We want to consider a
small spherical volume, of radius $r$, within the star that contains
$N$ electrons and $(A/Z) N$ nucleons.  After some algebra, the
potential energy $V_{\rm d} [r(x)]$ due to degeneracy pressure can be
written in the form 
\be
V_{\rm d} = N m_e {3 \over 8 x^3} \Bigl[ \Bigl\{ x (1 + x^2)^{1/2} 
( 1 + 2 x^2) - \ln [ x + (1 + x^2)^{1/2} ] \Bigr\} - 1 \Bigr] 
\, \equiv N m_e \Phi(x) \, , 
\ee
where the second equality defines a dimensionless function $\Phi(x)$. 
In the limit of small $x$ (low density), this potential has the usual 
form associated with nonrelativistic degenerate fermions, i.e., 
$\Phi \propto x^2 \propto r^{-2}$. In the opposite (relativistic) 
limit of large $x$, $\Phi \propto x \propto r^{-1}$. 
We also must include an additional contribution to the potential 
energy due to gravity. This potential can be represented with the 
approximate form 
\be
V_{\rm g} \approx - f {G M_N^2 \over r - r_S} \, , 
\ee
where $M_N = A N m_P$ is the total mass of the volume, 
$r_S = 2 G M_N$ is the corresponding Schwarzschild radius, 
and we have defined a geometrical factor $f \approx 3/5$.
Unless the radius $r$ approaches the Schwarzschild radius 
$r_S$, the degeneracy part of the potential completely dominates that
of gravity.  As a result, to a good approximation, we only need to
include $V_{\rm d}$ in the action integral [\ref{eq:action}], i.e., 
\be
S_T \approx 2 (9 \pi/4)^{1/3} [2 (A/Z) m_P/m_e ]^{1/2} \, N^{4/3} 
\, \int_{x_0}^\Omega \,  {dx \over x^2} \Phi^{1/2} (x) \, . 
\ee
The integration ranges from starting values $x_0$ near unity, 
corresponding to the initial density of the volume, to a large 
upper cutoff $\Omega \gg 1$, corresponding to black-hole-like 
densities. The exact value of the upper cutoff does not matter in 
practice because the integral has almost all of its support for 
small $x$ (low density). 

For white dwarfs, with $A/Z \approx 2$, we have 
\be
S_T \approx 335 \, N^{4/3} \, I (x_0) \, ,  
\label{eq:stunnel} 
\ee
where $I(x_0)$ is the integral evaluated for a given starting 
value of $x_0$, i.e., a given starting density.  The fastest tunneling 
rate will occur in the center of the star where the density (and $x_0$) 
is maximal. For a typical white dwarf of mass 0.5 $M_\odot$, 
the central density is about $10^6$ g/cm$^3$, $x_0 \approx$ 0.80, 
and $I(x_0) \approx$ 1.56. For a more massive white dwarf of 
1.0 $M_\odot$, the central density is about $3 \times 10^7$ g/cm$^3$, 
$x_0 \approx$ 2.5, and $I(x_0) \approx$ 1.0. 

For any given number $N$ of particles that are required to form a
black hole, equation [\ref{eq:stunnel}] determines the tunneling
probability $\propto \exp[-S_T]$. We can thus estimate the probability
(per natural oscillation time $\tau_0$) for white dwarfs to produce
black holes through a tunneling process, 
\be 
P_N = {\rm e}^{-335 I(x_0) N^{4/3}} \, , 
\ee
where $N$ determines the mass of the black hole 
($\mbh \approx (A/Z) N m_P$). This process is exponentially suppressed
as the number $N$ of particles increases, so the process is dominated
by the minimum value of $N$.

As a reference point, if the minimum mass for a black hole is given 
by the Planck scale, $\mpl$ $\sim 2 \times 10^{-5}$ g, we obtain 
$N \sim 10^{19}$ and hence $S_T \approx 10^{28}$.  For this incredibly
large value for the action, the natural vibration time scale $\tau_0$
becomes completely irrelevant.

The fastest possible rate for this process would be for the case $N=2$, 
and the action would be $S_T \approx$ 844 $I(x_0)$.  In the center of 
a relatively massive white dwarf, where $I \sim 1$, the 
corresponding time scale is $\tau \sim {\rm e}^{844} m_P^{-1} 
\sim 10^{336}$ yr.  This time scale is still long compared to 
that of gravitationally induced proton decay [1, 14--16, 18]. 
We thus conclude that the tunneling of aggregates of baryons into
black holes will not be an important process within white dwarfs.

The result is similar for the case of neutron stars [10].  In this case, 
we must replace the electron mass $m_e$ with the nucleon mass in all of 
the formulae and use $A/Z = 1$. The resulting action is much smaller,  
\be 
S_T \approx 5.4 N^{4/3} I(x_0) \, . 
\ee 
At the center of a neutron star, the density is $\sim 10^{15}$
g/cm$^3$ and hence approaches nuclear denisty. As a result, $x_0
\approx 1$, and hence $I(x_0)$ $\approx$ 1.4. The action then becomes
$S_T \approx 7.6 N^{4/3}$.  For the fastest case of only two nucleons,
$N=2$, we have $S_T$ $\approx$ 19.  This small action would seemingly
imply a reasonably high probability of tunneling into a black hole.
However, this calculation only takes into account the tunneling of
neutrons through the potential barrier provided by their degeneracy
pressure.  For small values of $N$ and high densities, the neutrons
start to overlap, and the quark structure of the nucleons must also be
taken into account.  The process of small numbers of quarks tunneling
into a black hole is one allowed channel for gravitationally induced
proton decay and the estimated time scale is about $10^{45}$ yr (see
[14, 18] and also the following discussion).

\bigskip 
\noindent
{\bf B. Kinematic Stellar Model} 

In this approximation, we consider the star to be a collection of
particles of mass $m$ with a number density $n$. We want to calculate
the probability that $N$ particles happen to lie within their own 
Schwarzschild radius, i.e., that of a black hole of mass $\mbh = m N$.  
In natural units, the volume of the hypothetical black hole is given by 
\be
V_{\rm bh} = 8 m^3 N^3 \mpl^{-6} \, . 
\ee
These $N$ particles generally occupy a much larger volume $V_0$ given by 
\be 
V_0 = N/n \, . 
\ee
Thus, the probability $p_1$ that a single particle lies within 
the black hole volume is simply 
\be 
p_1 = {V_{\rm bh} \over V_0} = 8 n m^3 N^2 \mpl^{-6} \, , 
\ee
and the probability $p_N$ that $N$ particles lie within the black 
hole volume is 
\be 
p_N = (p_1)^N = (8 n m^3 N^2 \mpl^{-6})^N  \, . 
\ee
Since the star contains $N_\star/N$ volumes of $N$ particles, the total 
probability $P_N$ that the star contains a black hole composed of 
$N$ particles is given by 
\be 
P_N = N_\star N^{-1} (8 n m^3 N^2 \mpl^{-6})^N  \, = 
N_\star \lambda^N N^{2N - 1} \, .
\ee
In the second equality, we have defined $\lambda$ = $8 n m^3 \mpl^{-6}$ 
$\approx 10^{-115}$ $\ll$ 1, where the numerical value assumes a nuclear 
density to evaluate $n$.  For a white dwarf density, $n$ will be far
smaller, and the value of $\lambda$ will decrease accordingly. 
For all values of $N$ less than $N_\ast$, the probability is a
decreasing function of $N$. In other words, the most probable black
hole to be formed kinematically will have the smallest possible value
of $N$.

For the case in which the minimum black hole mass is the Planck mass, 
$N$ $\approx$ $6 \times 10^{18}$, and hence $- \ln [P_N]$ $\sim$ 
$10^{21}$.  Loosely speaking, this time scale is somewhat 
longer than, but comparable to, the tunneling time scale calculated 
previously.  On the other hand, for small values of $N$, the time 
scale is ${\cal O} (\lambda^{-1} n^{1/3})$ $\sim$ $10^{89}$ yr. 
For small values of $N$, one can also construct a kinematic model 
of the proton as a system of $N=3$ quarks and make a similar calculation 
to model proton decay [18]. 

\bigskip 
\noindent
{\bf C. Virtual Black Holes} 

In any quantum theory, one expects to find vacuum fluctuations
associated with the fundamental excitations of the theory.  Thus, in
electrodynamics, one has the possibility of forming electron-positron
pairs for a short time directly out of the vacuum.  The existence of
such processes can be observed indirectly by many quantum phenomena,
such as the Casmir effect.  However, this example illustrates only 
the simplest such possibility.  One must also include in the vacuum
processes all excitations of the theory.  One should thus include the
possibility of the production of proton-antiproton pairs, or even
monopole-antimonopole pairs.  These processes will generally be highly
suppressed relative to the electron-positron amplitudes by virtue of
their correspondingly large masses.

In gravitation, one therefore expects not only to find virtual
gravitons playing a role, but also virtual black holes. Unfortunately, 
however, the theory of gravitation is unrenormalizable.  Although this
difficulty prevents one from doing reliable calculations, but it is
not unreasonable to suppose that a semi-classical calculation will
give reasonable answers.

Einstein gravity is controlled by the action [19] 
\be 
I[g] = - {1 \over 16 \pi G} \int_M R \, (-g)^{1/2} d^4x - 
{1 \over 8 \pi G} \int_{\partial M} K \, (-h)^{1/2} C[h] d^3x \, ,
\ee 
where $G$ is Newton's constant and $R$ is the Ricci scalar 
for the metric $g_{ab}$, which is defined on the spacetime $M$. 
The spacetime boundary $\partial M$ has the induced metric $h_{ab}$. 
The quantity $K$ is the trace of the second fundamental form 
on the boundary $\partial M$ and $C[h]$ is a functional 
of $h$ defined so that the action of Minkowski space vanishes.
Extremization of this action for fixed metric on the boundary 
leads to the Einstein equations for $g_{ab}$ in $M$. 

The path integral for gravity is
\be
Z \sim \int {\cal D}[g] \, {\rm e}^{i I[g]} \, , 
\ee
where the integral is taken over all metrics $g$. Our goal is 
to investigate how black holes contribute to any amplitude in 
quantum gravity. We first assume that this integral can be 
approximated by the usual Euclidean continuation [6, 13, 19]. 
The action for a single black hole of mass $m$ is then given by 
\be 
I_1 = {4 \pi m^2 \over \mpl^2} \, , 
\ee
where $\mpl$ is the Planck mass. Ignoring interactions between 
the black holes, we find the action for a collection of $N$ 
black holes to be 
\be 
I_N = {4 \pi N m^2 \over \mpl^2}  \, .
\ee
In the path integral, the black holes are indistinguishable; 
each is independent of the others and can be positioned anywhere 
in space.  Since $N$ is undetermined, we can evaluate $Z$ in a 
box of volume $V$ to obtain the result 
\be 
Z \sim \int_0^\infty dm \sum_{N=0}^\infty 
\exp[-4 \pi N m^2 / \mpl^2] \, {1 \over N!} \, 
\Bigl[ {V \over \ell_{\rm pl}^3} \Bigr]^N \, .
\ee
The factor of $V$ comes from accounting for the black holes 
being anywhere in the box, and the factor of $1/N!$ arises 
from their indistinguishability. 

The combination of these results thus defines a probability 
distribution for having $N$ black holes with mass $m$. Elementary 
calculations yield the corresponding expectation values for the 
number density of black holes and for the black hole mass, i.e., 
\be 
\langle n \rangle \sim {V \over \ell_{\rm pl}^3} 
\qquad {\rm and} \qquad \langle m \rangle \sim \mpl \, , 
\ee
where $\ell_{\rm pl}$ is the Planck length.  Thus, spacetime must be
filled with tiny Planck mass black holes with a density of roughly one
per Planck volume.  These microscopic virtual black holes will live 
roughly for one Planck time.  This picture of the spacetime vacuum is
sometimes called the spacetime foam.

Since black holes do not conserve baryon number, these virtual black
holes contribute to the rate of proton decay due to the gravitational
interaction. A proton can be considered to be a hollow sphere of
radius $10^{-13}$ cm that contains three (valence) quarks. Suppose
that two of these quarks fall into the same black hole at the same
time.  Since the black hole will decay predominantly into the lightest
particles consistent with the conservation of charge and angular
momentum, this process effectively converts the quarks into lighter
particles.  These particles will usually be electrons and neutrinos  
and hence baryon number conservation is generally violated. In other 
words, quantum gravity introduces an effective interaction leading to
processes of the form 
\be 
q + q \to \ell + \nu \, . 
\label{eq:interaction} 
\ee
This interaction can be regarding as a four-Fermi interaction 
whose coupling strength is determined by the Planck mass.  This 
process is mediated by black holes and can violate conservation 
of baryon number. (Note that this process cannot be mediated by 
just gravitons because such interactions conserve both electric 
charge and baryon number.) 

The probability of two quarks being within one Planck length
($\ell_{\rm pl} \sim 10^{-33}$ cm) of each other inside a proton is
about $(m_P/\mpl)^3 \sim 10^{-57}$. This value represents the
probability per proton crossing time $\tau_P \sim m_P^{-1} \sim
10^{-31}$ yr, if we assume that the particles move at the speed of
light.  In order for an interaction to take place (such as
[\ref{eq:interaction}]), a virtual black holes must be present at the
same time that the two quarks are sufficiently near each other.
Including this effect reduces the overall interaction probability by
an additional factor of $m_P/\mpl$.  Converting these results into a
time scale for proton decay, we find an estimated proton lifetime of
$\tau_P \sim 10^{45}$ yr.  Not surprisingly, this lifetime is also
what one would expect from equation [\ref{eq:gut}] if the unification
mass $M_X$ is taken to be the Planck mass $\mpl$.

\bigskip 
\noindent
{\bf D. Time Scales for Evaporation vs Accretion} 

Once a black hole exists within a star, two very different fates are
possible.  If the black hole is large enough, it will live long enough
that it accretes additional material from the star before evaporating
away.  In this case, the black hole can eventually accrete the entire
star.  On the other hand, if the black hole is sufficiently small, it
will evaporate before interacting with the stellar material.  In this
case, the legacy of the black hole is to leave behind its Hawking
radiation products.  Since black holes are known to not conserve
baryon number in their evaporation processes [6, 8, 13], the net 
result of this latter process is a mechanism for baryon decay. 

In order to determine the mass of a black hole required to survive, 
rather than evaporate, we set the Hawking evaporation time equal to 
the interaction time (both time scales depend on the mass $\mbh$ of 
the black hole). The evaporation time is given by 
\be
\tau_{\rm evap} = \tau_{\rm pl} (\mbh/\mpl)^3 \, , 
\ee
where $\tau_{\rm pl}$ is the Planck time. The interaction time is 
given by 
\be
\tau_{\rm int} = (n \sigma v)^{-1} = {\tau_{\rm pl} \over 4 \pi } 
(\mbh/\mpl)^{-2} (\mpl/\Lambda)^3 \, , 
\ee
where we have introduced an energy scale $\Lambda$ which defines 
the number density of the star through the relation $n = \Lambda^3$ 
(notice that $\Lambda \sim 200$ MeV for a neutron star and 
$\Lambda \sim 0.1$ MeV for a white dwarf). Notice also we have taken 
$v=c=1$, which results in the shortest possible interaction time. 
The critical mass black hole, the smallest black hole that will 
survive rather than evaporate, is thus given by 
\be
\mbhc = \mpl \, (4 \pi)^{-1/5} (\mpl/\Lambda)^{3/5} \, . 
\label{eq:bhcrit} 
\ee
For nuclear densities, such as those encountered in a neutron star, 
the critical mass black hole is about $\mbhc = 4 \times 10^{11} \mpl$ 
$\approx$ $4 \times 10^6$ g (about the mass of a tractor trailer). 
For the density of a white dwarf, the critical mass is somewhat 
larger, $\mbhc = 4 \times 10^{13} \mpl$. 

The corresponding values of $N$, the number of particles required to 
make the black hole, are also large: $N \approx 4 \times 10^{30}$ for 
the neutron star case and $N \approx 4 \times 10^{32}$ for the white dwarf. 

All of the mechanisms for black hole formation favor small black holes
over large ones.  In particular, the most probable black holes will be
much smaller than those required to survive and accrete the entire
star, rather than evaporate.  Even if the minimum possible mass for a
black hole is as large as the Planck mass $\mpl$, almost all black
holes will still evaporate long before they can interact and grow. 
The net result of black hole processes within degenerate stars is 
thus to create a channel for the evaporation of mass energy.  In
other words, black hole processes lead to the decay of baryons and the
loss of energy from the star.

\newpage 
\bigskip 
\centerline{\bf III. LONG TERM STELLAR EVOLUTION} 
\bigskip 

Given the results obtained above, the long term evolution of cold
degenerate stars is now clear: They will slowly evaporate.  

The rate at which the stars evaporate depends on the black hole
formation rate, which in turn depends on the minimum mass black hole
that can form.  If the smallest black hole mass is less than or
comparable to the proton mass, then the net result will be
gravitationally driven proton decay at a rate $\gp$.  If the
minimum black hole mass is larger and contains the mass equivalent of
$N$ nucleons (e.g., if the Planck mass is the minimum black hole mass,
$N$ $\approx$ $10^{19}$), then we can still define an effective
nucleon decay rate through the relation
\be 
\gp = N \Gamma_{\rm bh} \, , 
\ee
where $\gp$ is the rate of black hole production. This form 
does not apply when the black hole mass $\mbh = N m_P$ is 
larger than the critical mass $\mbhc$ required to survive 
rather than evaporate (see equation [\ref{eq:bhcrit}]). 

Once the decay rate for its constituent nucleons is determined, 
the star will evolve is a simple manner [1, 3]. The total 
luminosity is given by 
\be 
L_\star = \gp M_\star (t) = \gp M_{\star 0} {\rm e}^{-\gp t} \, , 
\label{eq:lumstar}
\ee
where $M_\star$ is the stellar mass and $M_{\star 0}$ is its 
initial mass. 

The decay products from black hole evaporation can be divided 
into two general classes: those that interact with stellar material 
before leaving the star (optically thick components) and those that 
leave with out interacting (optically thin components).  Photons 
and all charged particles will be optically thick, whereas neutrinos 
and gravitons will be optically thin. The fraction ${\cal F}$ of 
optically thick components will generally thermalize within the 
stellar interior and ultimately produce a photon luminosity 
\be 
L_{\star \gamma} = {\cal F} L_\star \, = 
4 \pi R_\star^2 \sigma_B T_\star^4 \, . 
\label{eq:lstartstar}
\ee
In the second equality we have defined the corresponding surface
temperature of the star through the usual black body relation, 
where $\sigma_B$ is the Stefan-Boltzmann constant. The stellar radius 
$R_\star$ is determined by the usual mass-radius relation for cold 
degenerate stars [7, 9], 
\be 
R_\star = 1.42 \Bigl( {\mpl \over m_D} \Bigr) \Bigl( {\mpl \over m_P} \Bigr)
\Bigl( {M_\star \over m_P} \Bigr)^{-1/3} m_P^{-1} \, , 
\label{eq:massrad}
\ee
where $m_D$ is the mass of the particle responsible for the 
degeneracy pressure, i.e., $m_D$ is the electron mass $m_e$ for 
white dwarfs and the neutron mass $m_N$ for neutron stars. 

Combining equations [\ref{eq:lstartstar}] and [\ref{eq:massrad}], 
we obtain an expression for the evolutionary tracks of the stars in 
the Hertzsprung-Russell (H-R) diagram, 
\be 
L_{\star \gamma} \approx 1.5 \, (4 \pi \sigma_B)^{3/5} \, 
({\cal F} \gp)^{2/5} \, \Bigl( {\mpl^2 \over m_D m_P} \Bigr)^{6/5} \, 
\Bigl( {M_\star \over m_P} \Bigr)^{-2/5} \, m_P^{-4/5} \, 
T_\star^{12/5} \, .
\label{eq:hrtrack} 
\ee 
In this regime, stellar evolution is thus determined once the
effective nucleon decay rate $\gp$ is specified. Unfortunately, as
discussed above, the decay rate $\gp$ can take a rather wide range of
values.  The resulting evolution is shown in Figure 1 for the 
representative value of $\gp$ = ($10^{45}$ yr)$^{-1}$.

Degenerate objects become larger in radial size as they lose mass. 
White dwarfs will follow the evolutionary tracks given by equation 
[\ref{eq:hrtrack}] until they expand to a size for which they are 
no longer degenerate.  At this point in the star's future evolution, 
when the stellar mass has fallen to $\sim 10^{-3}$ $M_\odot$, the 
star's track in the H-R diagram will change its slope. The above 
expressions for the stellar luminosity and the surface temperature
remain valid, but the mass-radius relation [\ref{eq:massrad}] no
longer applies.  The structure of non-degenerate matter is largely
determined by Coulomb forces, which enforce roughly uniform density
$\rho_0$, and the star follows an evolutionary track [1] given by 
\be
L_\star = {36 \pi \sigma_B^3 \over {\cal F}^2 \gp^2 \rho_0^2} 
\, T_\star^{12} \, ,  
\label{eq:hrtrack2} 
\ee
where $\rho_0 \sim 1$ g/cm$^3$.  At the start of this phase of
evolution, the object has a mass, radius, and density closely akin to
those of the planet Jupiter.  The object then continues to lose its mass
until it can no longer be considered a star.  The end point of stellar 
evolution occurs when the mass has fallen to about $10^{24}$ g, the 
point where the object can no longer completely thermalize its 
internal radiation.

Neutron stars follow a parallel evolutionary track given by equation
[\ref{eq:hrtrack}], where $m_D$ is now the neutron mass.  However, as
they lose mass and the neutrons are lifted out of degeneracy, neutron
stars must eventually experience a significant readjustment when the
mass drops below $\sim 0.0925 M_\odot$ [7].  During this
readjustment phase, a neutron star will either expand to a white dwarf
configuration [1] or explode and widely disperse its remaining mass
[20].  These possibilities are illustrated by the dashed curve and 
arrows shown in Figure 1. 

For completeness and comparison, we note that stellar black holes
evolve in a completely different manner. As black holes lose mass 
through Hawking evaporation, they become smaller while their surface 
temperature and luminosity increase.  Black holes follow an 
unambiguous track in the H-R diagram given by 
\be
L_{\rm bh} = {\sigma_B \over 4 \pi} T_{\rm bh}^2 \, . 
\label{eq:hrtrackbh} 
\ee
A portion of the black hole evolutionary path is shown in the lower
left part of Figure 1.  This evolution continues until the black hole
mass dwindles to the Planck mass and the surface temperature reaches
the Planck temperature.  Beyond this critical juncture, further black 
hole evolution depends on quantum gravity effects and is largely unknown. 

Notice that the long term evolution of white dwarfs is completely 
specified by equations [\ref{eq:hrtrack}] and [\ref{eq:hrtrack2}], 
and that all of the parameters are known except for the value of 
the effective proton decay rate $\gp$. For black holes, the long 
term evolution is given by equation [\ref{eq:hrtrackbh}] with 
absolutely no unknown parameters. 

\bigskip 
\centerline{\bf IV. SUMMARY} 
\bigskip  

The final demise of white dwarfs and neutron stars will take place on
a time scale that vastly exceeds the current age of the universe.
However, in the absence of conventional unification scale proton
decay, the black hole processes outlined here will determine the
ultimate fate of these cold degenerate stellar remnants. Within the
stars, the formation of small black holes is strongly favored over
large black holes and hence essentially all processes result in black
holes that evaporate rather than accrete additional stellar material.
The net result is that degenerate stars will themselves evaporate and
follow the evolutionary tracks depicted in Figure 1.  Since the vast
majority of stars are destined to end their conventional nuclear
burning lives as white dwarfs, with most of the remainder becoming
neutron stars, this work describes the long term fate and evolution of
almost all of the stars in the sky.

\bigskip 
\centerline{\bf Acknowledgements} 
\medskip 

We would like to thank G. Kane for facilitating the visit of M. Perry
to U. Michigan.  We also thank A. Zytkow for interesting discussions.
This work was supported by a Department of Energy grant, an NSF Young
Investigator Award, NASA Grant No.~NAG~5-2869, and by funds from the
Physics Department at the University of Michigan.

\bigskip 
\newpage 
\centerline{\bf REFERENCES} 
\medskip 

\medskip\par\pp{[1]} 
F. C. Adams and G. Laughlin, Rev. Mod. Phys. {\bf 69}, 337 (1997).  

\medskip\par\pp{[2]} 
P. Langacker, Phys. Rep. {\bf 72}, 186 (1984).   

\medskip\par\pp{[3]} 
D. A. Dicus, J. R. Letaw, D. C. Teplitz, and V. L. Teplitz, ApJ, 
{\bf 252}, 1 (1982).

\medskip\par\pp{[4]} 
G. Feinberg, Phys. Rev. D {\bf 23}, 3075 (1981). 

\medskip\par\pp{[5]} 
S. W. Hawking, Comm. Math. Phys. {\bf 43}, 199 (1975);  
S. W. Hawking, Nature {\bf 248}, 30 (1974). 

\medskip\par\pp{[6]} 
N. D. Birrell and P. C. W. Davies, {\sl Quantum Fields in Curved Space} 
(Cambridge Univ. Press, Cambridge, 1982). 

\medskip\par\pp{[7]} 
S. L. Shapiro and S. A. Teukolsky, {\sl Black Holes, White Dwarfs, 
and Neutron Stars} (Wiley, New York, 1983). 

\medskip\par\pp{[8]} 
K. S. Thorne, R. H. Price, and D. A. MacDonald, {\sl Black Holes: 
The Membrane Paradigm} (Yale Univ. Press, New Haven, 1986). 

\medskip\par\pp{[9]} 
S. Chandrasekhar, {\sl Stellar Structure} (Dover, New York, 1939).  

\medskip\par\pp{[10]} 
F. J. Dyson, Rev. Mod. Phys. {\bf 51}, 447 (1979). 

\medskip\par\pp{[11]} 
J. Dimock and B. S. Kay, Annales Inst. H. Poincare, A {\bf 37}, 93 (1982); 
J. Dimock and B. S. Kay, Annals Phys. {\bf 175}, 366 (1987).

\medskip\par\pp{[12]} 
I. D. Novikov and V. P. Frolov, {\sl Physics of Black Holes}
(Kluwer, Dordrecht, 1989). 

\medskip\par\pp{[13]} 
R. M. Wald, {\sl Quantum Field Theory in Curved Spacetime and 
Black Hole Thermodynamics} (Univ. Chicago Press, Chicago, 1994). 

\medskip\par\pp{[14]} 
Ya. B. Zeldovich, Phys. Lett. {\bf 59 A}, 254 (1976);  
Ya. B. Zeldovich, Sov. Phys. JETP, {\bf 45}, 9 (1977). 

\medskip\par\pp{[15]} 
S. W. Hawking, D. N. Page, and C. N. Pope, Phys. Lett. {\bf 86 B}, 
175 (1979). 

\medskip\par\pp{[16]} 
D. N. Page, Phys. Lett. {\bf 95 B}, 244 (1980). 

\medskip\par\pp{[17]} 
J. Ellis, J. S. Hagelin, D. V. Nanopoulos, and K. Tamvakis, 
Phys. Lett. {\bf 124 B}, 484 (1983). 

\medskip\par\pp{[18]} 
M. J. Perry, in {\sl Unification of Elementary Forces and Gauge Theories}, 
ed. D. B. Cline and F. E. Mills, p. 485 (Harwood, London, 1977). 

\medskip\par\pp{[19]} 
G. W. Gibbons and S. W. Hawking, Phys. Rev. D {\bf 15}, 2752 (1977). 

\medskip\par\pp{[20]} 
D. N. Page, Phys. Lett. {\bf 91 A}, 201 (1982).

\bigskip 
\bigskip 
\centerline{\bf Figure Caption} 
\medskip 

\noindent 
Figure 1.  The long term evolution of cold degenerate stars in the H-R
diagram. Upon completing the early stages of stellar evolution, white
dwarfs and neutron stars rapidly cool to an equilibrium temperature
dictated by black hole induced proton decay (assumed here to occur at
rate $\gp = [10^{45} \rm{yr}]^{-1}$).  The white dwarf models are
plotted at successive twofold decrements in mass.  The mean stellar
density (in Log[$\rho$/g]) is indicated by the gray scale shading,
and the sizes of the circles are proportional to stellar radius. The
relative size of the Earth and its position on the diagram are shown
for comparison.  The evaporation of a 1 $M_{\odot}$ neutron star is
illustrated by the parallel sequence, which shows the apparent radial
sizes greatly magnified for clarity. The Hawking radiation sequence
for black holes is also plotted.  The arrows indicate the direction of
time evolution.

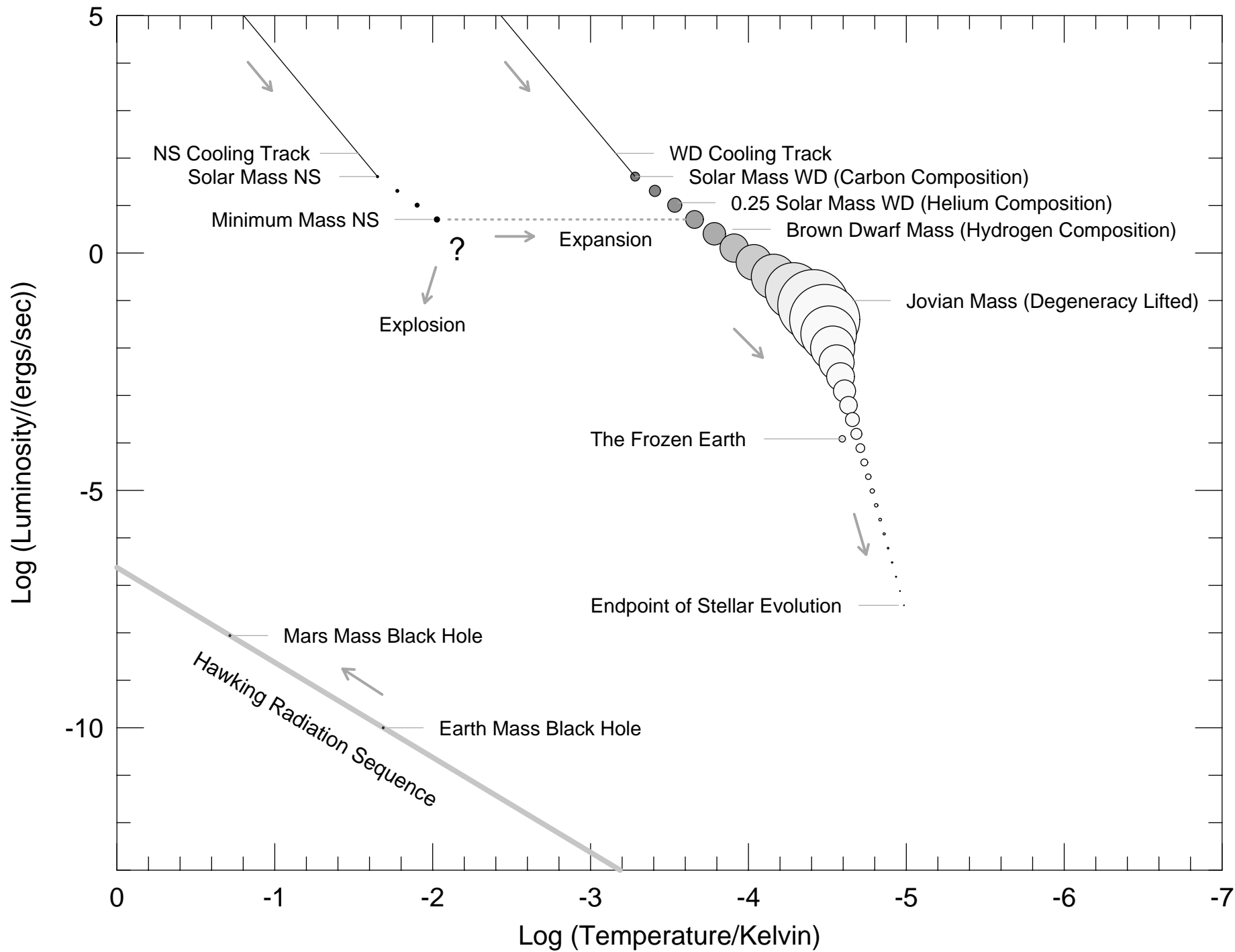
\end{document}